\documentclass[preprint,12pt]{elsarticle}

 \usepackage{graphicx}

\usepackage{amssymb}
\usepackage{amsmath}

\usepackage[nodots]{numcompress}

\journal{Applied Ocean Research}

\begin{document}

\begin{frontmatter}



\title{Relations for a periodic array of flap-type wave energy converters}


\author[ucd]{E. Renzi\corref{cor}}
\ead{emiliano.renzi@ucd.ie}

\author[ucd,ca]{F. Dias}
\ead{frederic.dias@ucd.ie}

\cortext[cor]{Corresponding author}

\address[ucd]{UCD School of Mathematical Sciences, University College Dublin, Belfield Dublin 4, Ireland}
\address[ca]{Centre de Math\'{e}matiques et de Leurs Applications (CMLA), Ecole Normale Sup\'{e}rieure de
Cachan, 94235 Cachan, France}

\begin{abstract}
This paper investigates the interaction of plane incident waves with a wave farm in the open ocean. The farm consists of a periodic array of large flap-type wave energy converters. A linear inviscid potential-flow model, already developed by the authors for a single flap in a channel, is considered. Asymptotic analysis of the wave field allows to obtain new expressions of the reflection, transmission and radiation coefficients of the system. It is shown that, unlike a line of heaving buoys, an array of flap-type converters is able to exploit resonance of the system transverse modes in order to attain high capture factor levels. Relations between the hydrodynamic coefficients are derived and applied for optimising the power output of the wave farm.
\end{abstract}


\end{frontmatter}


\section{Introduction}
\label{sec:intro}
Research on wave energy converters (WECs) has concentrated traditionally on systems of small floating bodies, like offshore heaving buoys (see \cite{EV76}--\cite{ FA02}). However, the seminal theories on WECs that originated from this first scientific approach to wave energy extraction in the 1970s, do not capture exhaustively the dynamics of the last-generation WECs. The latter are usually large-scale devices designed to be deployed in arrays, some of them in the near-shore environment. For example, while studying the dynamics of an offshore heaving WEC in a channel, Srokosz (1980) \cite{SR80} showed that resonance of the channel sloshing modes is detrimental to the efficiency of power absorption. Conversely, in a recent analysis of a large flap-type WEC in a channel, Renzi \& Dias \cite{RD12} noted that the trapping of transverse modes near the flap increases the efficiency of the converter. Because of the image effect of the channel walls, this fact is also expected to occur in an infinite array of flap-type converters. The aim of this work is to discover the dynamics of a system of last-generation flap-type WECs and to outline its similarities and differences with respect to the systems of the first generation. As a result of this analysis, an optimisation criterion for an array of flap-type WECs is devised, depending on the physical and geometrical parameters of the system.

In Section \ref{sec:model} the behaviour of an array of converters in the open ocean is investigated by taking as a reference the theoretical framework of Renzi and Dias \cite{RD12}. The expressions of the free-surface elevation for the diffracted and radiated wave field in the fluid domain are derived accordingly. Analysis of the wave motion in the far field allows to obtain new formulae for the reflection, transmission and radiation coefficients. Various relations between the hydrodynamic coefficients are then shown in Section \ref{sec:rel}. Some of these relations correspond directly to Srokosz's results \cite{SR80} for floating bodies of symmetric shape in a channel. Some others, on the other hand, incorporate specific properties of the wave field generated by the flap-type converter, not considered before, and point out the peculiarity of such WEC with respect to the converters of the first generation. The analytical model is validated against known theories in the small-gap and in the point-absorber limit. In Section \ref{sec:paran}, a parametric analysis is undertaken for optimising the performance of the system. It is shown that the maximum capture factor is attained at complete trapping of the transverse modes of the array. When complete trapping is not possible, partial trapping can still increase the performance of the system. Finally, in Section \ref{sec:waven} a practical application of an array of large flap-type WECs is devised. Comparison with available data obtained with a finite-element numerical code is very satisfactory (see \ref{sec:appD}).
\section{Mathematical model}
\label{sec:model}
\subsection{Theoretical background}
\begin{figure}
  \centerline{\includegraphics[width=9cm, trim=3.5cm 0cm 3.5cm 0cm]{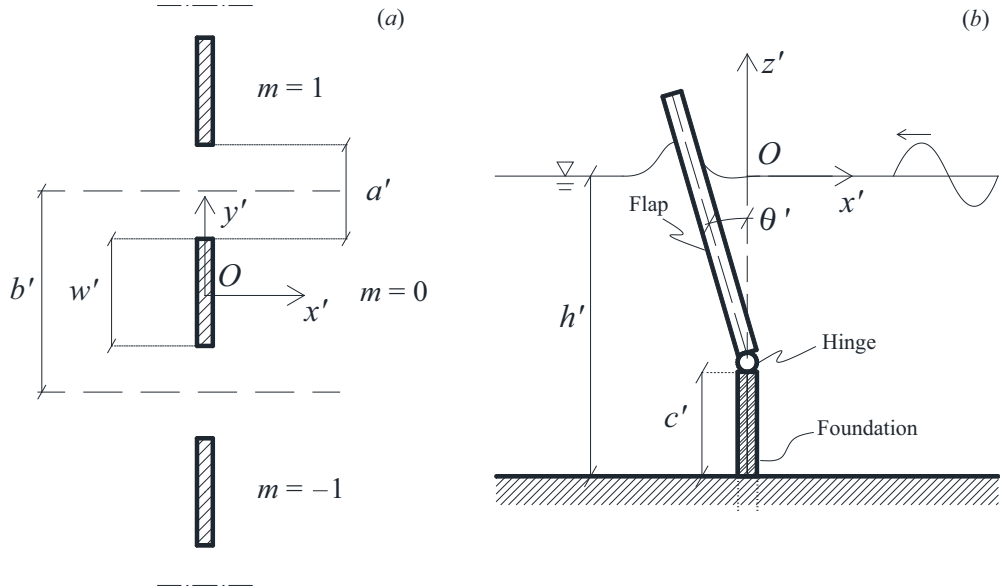}}
  \caption{Geometry of the array ($a$) and the reference flap ($b$) in physical variables.}
\label{fig:geom}
\end{figure}

Consider an in-line array of identical flap-type wave energy converters, each hinged on a bottom foundation of height $c'$ in an ocean of constant depth $h'$, as shown in figure \ref{fig:geom}. 
Primes denote dimensional quantities. Monochromatic incident waves of amplitude $A_I'$, period $T'$ and frequency $\omega'=2\pi/T'$ are incoming from the right and set the flaps into motion, which is converted into useful energy by means of generators linked to each device. Since the practical applications of such a system are usually in the nearshore \citep{WF12}, where wave fronts are almost parallel to the shoreline because of refraction, normal incidence is also assumed. Let $w'$ and $b'$ be the width of each flap and the spatial period of the array, respectively. Then the gap between two consecutive flaps is $a'=b'-w'$ (see again figure \ref{fig:geom}). A Cartesian coordinate system is set, with the $x'$ direction orthogonal to the flaps, the $y'$ axis along the flap lineup and the $z'$ axis rising from the undisturbed water level $z'=0$, positive upwards; $t'$ denotes time. Due to periodicity, the origin of the system can be set arbitrarily on any flap, which is therefore identified as the reference flap. The analysis is performed in the framework of a linear inviscid potential-flow theory for small-amplitude oscillations. The velocity potential $\Phi'$ must satisfy the Laplace equation 
\begin{equation}
\nabla'^2\Phi'(x',y',z',t') =0
\label{eq:lapl}
\end{equation}
in the fluid domain. The linearised kinematic-dynamic boundary condition on the free surface reads
\begin{equation}
\Phi'_{,t't'}+g \Phi'_{,z'}=0, \quad z'=0,
\label{eq:bcsurf}
\end{equation}
where $g$ is the acceleration due to gravity and subscripts with commas denote differentiation with respect to the relevant variables. Absence of normal flux at the bottom yields
\begin{equation}
\Phi'_{,z'}=0, \quad z'=-h'.
\label{eq:bottom}
\end{equation}
Because of normal incidence of the incoming wave field, periodicity of the problem requires 
\begin{equation}
f'(x',y'+mb',z',t')=f'(x',y',z',t'),\quad m=0,\pm 1,\pm 2,\dots,\; y'\in(-b'/2,b'/2),
\label{eq:period}
\end{equation}
where $f'$ indicates any physical quantity associated to the problem and $m$ each of the flaps; $m=0$ denotes the reference flap. Extension to an obliquely-incident wave field can be easily made \citep[see for example][]{PE96}. However, since flap-type WECs are usually designed to operate under normally-incident waves \citep[see][]{WF12}, only normal incidence will be considered here. Because of the periodicity condition (\ref{eq:period}), the solution to the complete problem can be obtained by investigating the wave interaction with the reference flap centred at the origin, with $|y'|<b'/2$. Symmetry of the problem requires also
\begin{equation}
\Phi'_{,y'}=0,\quad y'=\pm b'/2,
\label{eq:simm}
\end{equation}
which can be regarded as a no-flux boundary condition on two imaginary waveguides at $y'=\pm b'/2$ (see again figure \ref{fig:geom}). Let $\theta'(t')$ be the angle of rotation of the flap, positive if anticlockwise; then the kinematic boundary condition on the flap yields
\begin{equation}
\Phi'_{,x'}=-\theta'_{,t'}(t') (z'+h'-c') H(z'+h'-c'), \quad x'=\pm 0, |y'|<w'/2,
\label{eq:plate}
\end{equation}
where the thin-body approximation has been applied \citep{LM01}. The Heaviside step function in (\ref{eq:plate}) assures absence of flux through the bottom foundation. The problem defined above is formally equivalent to that solved by Renzi and Dias \cite{RD12} for a single converter in a channel. Here the main arguments of the theory in \cite{RD12} are retraced and applied to the array configuration. First, the system (\ref{eq:lapl})--(\ref{eq:plate}) is non-dimensionalised as follows \citep[see][eqn. (2.1)]{RD12}
\begin{eqnarray}
&(x,y,z,h,w,a,c)=(x',y',z',h',w',a',c')/b',\; t=\sqrt{g/b'}\,t',\nonumber \\
&\Phi=\left(\sqrt{gb'}A'\right)^{-1}\Phi',\; \theta=(b'/A') \theta',
\label{eq:nondimvar}
\end{eqnarray}
where the wave amplitude scale $A'\ll b'$ because of the hypothesis of small-amplitude oscillations. In expression (\ref{eq:nondimvar}), $a=(1-w)\in(0,1)$ defines the aperture of the array. Then time is factored out by setting
\begin{equation}
\Phi(x,y,z,t)=\Re\left\lbrace \phi(x,y,z) e^{-i\omega t}\right\rbrace,\quad \theta(t)=\Re\left\lbrace \Theta e^{-i\omega t}\right\rbrace,
\label{eq:separ}
\end{equation}
with $\omega=\sqrt{b'/g}\,\omega'$ \citep[see][eqn. (2.11)]{RD12}. The global spatial potential 
$$
\phi=\phi^R+\phi^S
$$
is the sum of the radiation potential $\phi^R$ and the scattering potential $\phi^S$. The latter is in turn decomposed into
$$\phi^S=\phi^I+\phi^D,$$ where
\begin{equation}
\phi^I(x,y,z)=-\frac{i A_I}{\omega\cosh kh}\cosh k(z+h) e^{-i kx}
\label{eq:incidentwav}
\end{equation}
is the incident wave potential and $\phi^D$ is the diffraction potential. $\phi^R$ and $\phi^D$ must be both outgoing at large $|x|$. In (\ref{eq:incidentwav}), $A_I=A'_I/A'$ is the non-dimensional amplitude of the incident wave and $k$ is the wavenumber, corresponding to the real solution of the dispersion relationship $\omega^2=k \tanh kh$.
Following the method described in Appendix B of \cite{RD12}, application of the Green integral theorem yields two hypersingular integral equations, in terms of the jump in radiation and scattering potentials across the plate \citep[see][eqn. (B10)]{RD12}. Those equations are solved by expanding the jumps in potential into series of Chebyshev polynomials of even order \citep[for details, see eqn.s (B11)--(B18) of][]{RD12}. Careful treatment of the singularity \citep[see][eqn. (B19)]{RD12} ultimately allows to write the potentials in the reference domain $|y|<1/2$ in a new semi-analytical form \citep[see][eqn.s (B24) and (B25)]{RD12}. The radiation potential is
\begin{equation}
\phi^R(x,y,z)=\sum_{n=0}^{+\infty}\sum_{p=0}^P\sum_{m=-\infty}^{+\infty}\phi^R_{npm}(x,y,z),
\label{eq:radpotsum}
\end{equation}
where
\begin{eqnarray}
\phi^R_{npm}(x,y,z)&=&-\frac{i w V}{8}\,\kappa_n x Z_n(z) \alpha_{(2p)n}\int_{-1}^{1} \left(1-u^2\right)^{1/2} \nonumber\\
&\times& U_{2p}(u)
\frac{H_1^{(1)}\left(\kappa_n\sqrt{x^2+(y-\frac{1}{2} wu-m)^2} \right)}{\sqrt{x^2+(y-\frac{1}{2} wu-m)^2}}\,du,
\label{eq:radpot}
\end{eqnarray}
$H_1^{(1)}$ being the Hankel function of the first kind and first order. In (\ref{eq:radpot}),  $V=i\omega\Theta$ is the complex angular velocity of the flap, the subscript $m$ identifies the contribution of each single flap, while the subscript $p$ indicates the order of the Chebyshev expansion, $U_{2p}$ being the Chebyshev polynomial of the second kind and even order $2p$, $p=0,1,...,P\in\mathbb{N}$. The subscript $n$ identifies the contribution of each depth mode
\begin{equation}
Z_n(z)=\frac{\sqrt{2}\cosh \kappa_n (z+h)}{\left(h+\omega^{-2}\sinh^2 \kappa_n h \right)^{1/2}},\quad n=0,1,\dots,
\label{eq:Zn}
\end{equation}
where $\kappa_0=k$, while $\kappa_n=ik_n$ denote the complex solutions of the dispersion relationship
\begin{equation}
\omega^2=-k_n\tan k_nh,\quad n=1,2,\dots
\label{eq:disprel}
\end{equation}
Finally, the $\alpha_{(2p)n}$ are the complex solutions of a system of linear equations ensuring that $\phi^R$ satisfies the kinematic condition on the flap \citep[see][eqn.s (B22) and (B23)]{RD12}. This system is solved numerically with a collocation scheme, therefore the solution (\ref{eq:radpot}) is partly numerical. In summary, $\phi_{npm}$ (\ref{eq:radpot}) indicates the $n$-th depth mode, $p$-th order potential of the wave field radiated by the $m$-th flap ($m$-th array mode), moving at unison with all the other flaps. The diffraction potential is given by
\begin{equation}
\phi^D(x,y,z)=\sum_{p=0}^P\sum_{m=-\infty}^{+\infty}\phi^D_{pm},
\label{eq:difpotsum}
\end{equation}
where
\begin{eqnarray}
\phi^D_{pm}(x,y,z)&=&-\frac{i w A_I}{8}\,kx\, Z_0(z)\beta_{2p}\int_{-1}^{1} \left(1-u^2\right)^{1/2} \nonumber\\
&\times& U_{2p}(u)
\frac{H_1^{(1)}\left(k\sqrt{x^2+(y-\frac{1}{2} wu-m)^2} \right)}{\sqrt{x^2+(y-\frac{1}{2} wu-m)^2}}\,du.
\label{eq:difpot}
\end{eqnarray}
In the latter, the $\beta_{2p}$ are the complex solutions of a system of linear equations, which ensures that $\phi^D$ satisfies the no-flux condition on the flap \citep[see][eqn.s (B22) and (B23)]{RD12}. Again, $\phi^D_{pm}$ indicates the $p$-th order potential diffracted by the $m$-th flap, in the presence of all the other flaps. Note that in $\phi^D$ (\ref{eq:difpot}) only the $0$-th depth mode is present, as required by the solvability of the whole radiation-diffraction problem \citep[see Appendix B.2 of][]{RD12}. Computational aspects involved in the numerical evaluation of (\ref{eq:radpot}) and (\ref{eq:difpot}) are detailed in \S 2.2 of \cite{RD12}.

\subsection{Body motion}
The equation of motion of the reference flap in the frequency domain is that of a damped harmonic oscillator \citep[see eqn. (2.33) of][]{RD12}, namely
\begin{equation}
\left[-\omega^2(I+\mu)+C-i\omega(\nu+\nu_{pto})\right]\Theta=F,
\label{eq:bodymot}
\end{equation}
depending on the moment of inertia of the flap $I=I'/(\rho b'^5)$, on the flap buoyancy torque $C=C'/(\rho g b'^4)$ and on the power take-off (PTO) coefficient $\nu_{pto}=\nu'_{pto}/(\rho b'^4\sqrt{gb'})$, where $\rho$ is the water density. The latter parameters are assumed to be all known. In (\ref{eq:bodymot})
\begin{equation}
\mu=\frac{\pi w}{2\sqrt{2}}\,\Re\left\lbrace \sum_{n=0}^{\infty} \alpha_{0n} \frac{\kappa_n(h-c)\sinh \kappa_n h+\cosh \kappa_n c-\cosh\kappa_n h}{\kappa_n^2\left(h+\omega^{-2}\sinh^2\kappa_n h \right)^{1/2}} \right\rbrace
\label{eq:mu}
\end{equation}
is the added inertia torque \citep[see eqn. (2.34) of][]{RD12}, while
\begin{equation}
\nu=\frac{\pi w}{2\sqrt{2}}\Im\left\lbrace \alpha_{00} \right\rbrace \frac{\omega\left[k(h-c)\sinh k h+\cosh k c-\cosh k h\right]}{k^2\left(h+\omega^{-2}\sinh^2 kh \right)^{1/2}} 
\label{eq:nu}
\end{equation}
and
\begin{equation}
F=-\frac{i\pi w A_I}{2\sqrt{2}}\beta_{0}\frac{\omega\left[k(h-c)\sinh kh+\cosh kc-\cosh kh\right]}{k^2\left(h+\omega^{-2}\sinh^2 kh\right)^{1/2}}
\label{eq:F}
\end{equation}
denote, respectively, the radiation damping \citep[see eqn. (2.35) of][]{RD12} and the complex exciting torque \citep[see eqn. (2.36) of][]{RD12}. If the PTO system is designed such that
$$\nu_{pto}=\sqrt{\frac{\left[C-(I+\mu)\omega^2\right]^2}{\omega^2}+\nu^2},$$
which corresponds to the optimum PTO damping \citep[see eqn. (2.40) of][]{RD12},
then the average generated power over a period is
\begin{equation}
P=\frac{1}{4}|F|^2\left[\sqrt{\frac{\left[C-(I+\mu)\omega^2\right]^2}{\omega^2}+\nu^2}+\nu\right]^{-1}.
\label{eq:pow}
\end{equation}
Now, the generated power (\ref{eq:pow}) is maximum under resonant amplification of the body motion, which occurs when
\begin{equation}
\omega=\sqrt{\frac{C}{I+\mu}}.
\label{eq:bodyres}
\end{equation}
By substitution of the latter expression into (\ref{eq:pow}), the optimum power available for extraction from each flap is therefore
\begin{equation}
P_{opt}=\frac{1}{8}\frac{|F|^2}{\nu},
\label{eq:Popt}
\end{equation}
which matches the well-known result of Srokosz \cite{SR80}.
The performance of each element of the array  is assessed quantitatively by using two main factors. The amplitude factor 
\begin{equation}
A_F=\frac{(h-c)\tan|\Theta|}{A_I}
\label{eq:AF}
\end{equation}
is defined as the ratio between the flap horizontal stroke and the amplitude of the incident waves, $\Theta$ being the solution of the equation of motion (\ref{eq:bodymot}). Finally, the capture factor is defined as the ratio between the power extracted per unit flap width and the power available per unit crest length
\begin{equation}
C_F=\frac{P}{\frac{1}{2} A_I^2 C_g w},
\label{eq:CF}
\end{equation}
where
\begin{equation}
C_g=\frac{\omega}{2k}\left(1+\frac{2kh}{\sinh 2kh}\right)
\end{equation}
is the group velocity of the incident waves. Since $P=P(a)$ and $w=1-a$, the capture factor (\ref{eq:CF}) depends intrinsically on the aperture of the array. A strength point of the method of \cite{RD12} is that knowing the coefficients $\alpha_{0n}$ and $\beta_0$ is sufficient to obtain immediately all the physical quantities describing the performance of the device (eqn.s \ref{eq:mu}--\ref{eq:CF}), without need to evaluate the potentials (\ref{eq:radpotsum}) and (\ref{eq:difpotsum}).
The wave motion at large distance from the array will be now analysed.
\section{The far field}
In this section the behaviour of the wave field is investigated at large distance from the array. First, consider the radiation potential $\phi^R_{npm}$ given by (\ref{eq:radpot}). For $n>0$ the Hankel function in (\ref{eq:radpot}) can be rewritten as
\begin{eqnarray}
H_1^{(1)}\left(\kappa_n|x|\sqrt{1+\left(\frac{y}{x}-\frac{wu}{2x}-\frac{m}{x}\right)^2}\right)\nonumber\\
=-\frac{2}{\pi}K_1\left(\kappa_n|x|\sqrt{1+\left(\frac{y}{x}-\frac{wu}{2x}-\frac{m}{x}\right)^2} \right),
\label{eq:modbes}
\end{eqnarray}
where $K_n$ denotes the modified Bessel function of the second kind and order $n$ \citep[see \S 8.407 of][]{GR07}. Since $K_1(z)\propto e^{-z}$ as $z\rightarrow\infty$ in (\ref{eq:modbes}) and hence in (\ref{eq:radpot}), the argument of $\phi^R_{npm}$ for $n>0$ decays exponentially in the far field, so that at leading order
\begin{equation}
\phi^R_{npm}\sim 0, \quad |x|\rightarrow\infty\,,n>0.
\end{equation} 
This happens since the modes $n>0$ physically represent the parasite waves generated by the motion of the flaps. These remain trapped near the device and do not contribute to the wave motion in the far field \cite{ME05}. As a consequence,
\begin{equation}
\phi^R\sim \sum_{p=0}^P\sum_{m=-\infty}^{+\infty}\phi^R_{0pm},\quad |x|\rightarrow\infty.
\label{eq:phirapp}
\end{equation}
Now substituting (\ref{eq:radpot}) into (\ref{eq:phirapp}), using the asymptotic expression (\ref{eq:Sff}) with $(X,Y)=(x,y-wu/2)$ and the integral formulae (\ref{eq:Ip0}), (\ref{eq:Ipq2}) and finally developing some straightforward algebra, yields
\begin{equation}
\phi^R\sim -\frac{iV}{\omega}\frac{\cosh k(z+h)}{\cosh kh}\,\sum_{q=0}^{\bar{q}}\mathcal{A}_q^{\pm}e^{\pm i\gamma_q kx} \cos(2q\pi y),\quad x\rightarrow\pm\infty.
\label{eq:phirff}
\end{equation}
In the latter expression, $\gamma_q=\sqrt{1-(2q\pi/k)^2}$ and $\bar{q}$ is the largest integer for which $\gamma_q$ is real, while
\begin{eqnarray}
\mathcal{A}_0^{\pm}&=&\mp\frac{i\pi}{8}\,w\,\omega\alpha_{00}Z_0(0),\label{eq:A0}\\
\mathcal{A}_q^{\pm}&=&\mp \frac{i}{4}\,w\,\omega Z_0(0)\, \epsilon_q\sum_{p=0}^{P} \alpha_{(2p)0}(-1)^p(2p+1)\frac{J_{2p+1}(q\pi w)}{qw}.\label{eq:Aq}
\end{eqnarray}
In (\ref{eq:Aq}), $\epsilon_q$ is the Jacobi symbol, while $J_{2p+1}$ is the Bessel function of first kind and order $2p+1$. Note that the radiation potential in the far field (\ref{eq:phirff}) is the sum of a progressive long-crested wave (term $q=0$) and several progressive short-crested waves (terms $0<q<\bar{q}$), which correspond to the propagating sloshing modes of the equivalent channel configuration of \cite{RD12}. Expression (\ref{eq:phirff}) is similar in form to (2.24) of \citep{SR80} (accounting for the various differences in the nomenclature), which gives the far-field expression of the radiation potential for a floating body, symmetric with respect to the $x$ axis, in a channel. In (2.24) of \cite{SR80}, however, the coefficients $\mathcal{A}_{q}^{\pm}$ are left in a general form, while here they are determined explicitly for the flap-type converter. The same steps can be repeated to find the far-field expression of the diffraction potential $\phi^D$ (\ref{eq:difpotsum}). By substituting (\ref{eq:difpot}), (\ref{eq:Sff}), (\ref{eq:Ip0}) and (\ref{eq:Ipq2}) in (\ref{eq:difpotsum}) and developing the algebra, the diffraction potential in the far field becomes
\begin{equation}
\phi^D(x,y)\sim \mp\frac{iA_I}{\omega}\frac{\cosh k(z+h)}{\cosh kh}\sum_{q=0}^{\bar{q}}R_q e^{\pm i\gamma_q kx}\cos(2q\pi y),\quad x\rightarrow\pm\infty,
\label{eq:phidff}
\end{equation}
where
\begin{eqnarray}
R_0&=&-\frac{i\pi}{8}\,w\,\omega\beta_0Z_0(0),\label{eq:R0}\\
R_q&=&-\frac{i}{4}\,w\,\omega Z_0(0)\,\epsilon_q\sum_{p=0}^P\beta_{2p}(-1)^p(2p+1)\frac{J_{2p+1}(q\pi w)}{qw}.\label{eq:Rq}
\end{eqnarray}
Equation (\ref{eq:phidff}) is similar to (2.25) of \cite{SR80}, in which, however, the $R_q$ are left in a general form.
Note that the calculation of the coefficients $\mathcal{A}_q^\pm$ and $R_q$ is straightforward once the linear system for $\alpha_{(2p)0}$ and $\beta_{2p}$ is solved \citep[see eqn. B23 of][]{RD12}.
\subsection{The free-surface elevation}
\begin{figure}[t]
\begin{center}
\includegraphics[width=8cm, trim= 5cm 9cm 5cm 0cm]{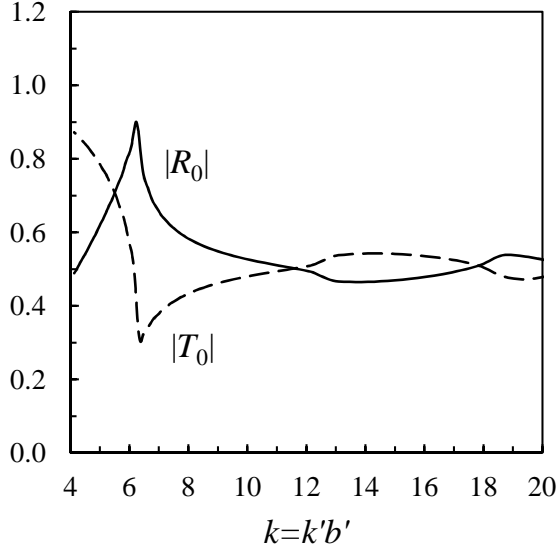}
\end{center}
\caption{Magnitude of the fundamental reflection and transmission coefficients, respectively $R_0$ (\ref{eq:R0}) and $T_0$ (\ref{eq:T0}), versus the non-dimensional wavenumber $k$. In this layout the flap width equals the gap size, i.e. $a=w=1/2$.
}
\label{fig:porevcom}
\end{figure}

The amplitude of the free surface in the far field is an important parameter in order to assess the impact of the array on the wave climate of the surrounding area. Given the total potential $\Phi(x,y,t)$, the free-surface elevation is
$$\zeta(x,y,t)=-\Phi_{,t}|_{z=0}=\Re\left\lbrace\eta(x,y) e^{-i\omega t}\right\rbrace,$$
where
\begin{equation}
\eta(x,y)=i\omega(\phi^I+\phi^R+\phi^D)
\end{equation}
is the relevant complex spatial component. Substituting (\ref{eq:incidentwav}) for $\phi^I$ and the far-field expressions (\ref{eq:phirff}) and (\ref{eq:phidff}) for $\phi^R$ and $\phi^D$, respectively, yields
\begin{equation}
\eta(x,y)\sim \left\lbrace
\begin{tabular}{l l}
$A_Ie^{-ikx}+\sum_{q=0}^{\bar{q}}(A_I R_q+V\mathcal{A}_q^+)e^{i\gamma_q kx}\cos(2q\pi y)$, & $x\rightarrow+\infty$\\
$\sum_{q=0}^{\bar{q}}(A_I T_q+V\mathcal{A}_q^-)e^{-i\gamma_q kx}\cos(2q\pi y)$,& $x\rightarrow-\infty$
\end{tabular}
\right.,
\label{eq:etaff}
\end{equation}
where
\begin{equation}
T_0=1-R_0, \quad T_q=-R_q.
\label{eq:T0}
\end{equation}
Overall, the free-surface elevation is the sum of a long-crested wave (term $q=0$) and several short-crested waves (terms $0<q<\bar{q}$), namely the propagating transverse modes of the array. Physically, in (\ref{eq:etaff}) the terms $\mathcal{A}_q^{\pm}$ represent the $q$th-mode radiation coefficients, $R_q$ is the $q$th-mode reflection coefficient and finally $T_q$ represents the $q$th-mode transmission coefficient. They enjoy all the general properties of the analogous terms introduced by Srokosz \cite{SR80} for bodies of symmetric shape in a channel. In addition, such coefficients have some specific properties, peculiar to flap-type bodies, which derive from their analytical structure, as shown in detail in \S\ref{sec:rel}. Figure \ref{fig:porevcom} shows the behaviour of $R_0$ and $T_0$ against the nondimensional wavenumber $k$ for a typical configuration where $a=1/2$. The plots in figure \ref{fig:porevcom} compare favourably with those of Williams \& Crull \cite[fig. 3]{WC93} and Porter \& Evans \cite[fig. 2]{PE96}, who studied the scattering of incident waves by an array of thin screens. 
Note the spiky behaviour of the coefficients, with spikes occurring at the resonant wavenumbers $k=2(\bar{q}+1) \pi$, $\bar{q}=0,1,\dots$  of the transverse short-crested waves, for which the $(\bar{q}+1)$th transverse mode turns from trapped to propagating.
In the following, the coefficients $R_0$, $T_0$ and $\mathcal{A}_0^\pm$ will be shown to enjoy some interesting properties and to be very useful for determining some relations between the hydrodynamic coefficients of the system.
\section{Derivation of relations for an array of flap-type WECs}
\label{sec:rel}
In this section, relations are derived for an array of flap-type WECs, based on the results obtained in the previous section. Some of these relations correspond directly to Srokosz's results \cite{SR80} for floating bodies of symmetric shape, while some others incorporate specific properties of the wave field (\ref{eq:etaff}) generated by the array of flap-type converters. In this sense, such expressions are new and point out the peculiarity of flap-type WECs with respect to the converters of the first generation. 
\subsection{Extended Bessho-Newman relation}
First consider $\mathcal{A}_0^+$ and $R_0$. From (\ref{eq:A0}) and (\ref{eq:R0}) it is immediate to get, respectively,
$\arg(\mathcal{A}_0^+)=\arg(\alpha_{00})-\pi/2$ and $\arg(R_0)=\arg(\beta_0)-\pi/2$. Since $\arg(\alpha_{00})=\arg(\beta_0)$ \citep[see Appendix C of][]{RD12}, then the complex coefficients $\mathcal{A}_0^+$ and $R_0$ must have the same argument, say $\delta$, for which
\begin{equation}
\mathcal{A}_0^+=|\mathcal{A}_0^+|e^{i\delta},\quad R_0=|R_0|e^{i\delta},
\label{eq:delta}
\end{equation}
for any wavenumber $k$. The same relation, but only for $k<2\pi$, can be also derived from the Bessho-Newman relation
\begin{equation}
A_0^+-\sum_{q=0}^{\bar{q}}\frac{\gamma_q}{\epsilon_q}\left(A_q^{+\ast}R_q+A_q^{-\ast}T_q \right)=0,
\label{eq:BN}
\end{equation}
where $()^\ast$ denotes the complex conjugate. Expression (\ref{eq:BN}) is obtained by applying Green's integral theorem to $\phi^S$ and $(\phi^R-\phi^{R\ast})$ and corresponds to (3.2) of \citep{SR80}, with small variations due to the difference in the nomenclature. Note that the Bessho-Newman relation (\ref{eq:BN}) is a general form valid for any floating body, symmetric with respect to the $x$ axis, in a channel (or for an infinite array of such bodies). Considering $k<2\pi$, i.e. $\bar{q}=0$, and using the identities $\mathcal{A}_0^-=-\mathcal{A}_0^+$ (see \ref{eq:A0}), with $\mathcal{A}_0^+=|\mathcal{A}_0^+|e^{i\delta}$, and $T_0=1-R_0$ (see \ref{eq:T0}), (\ref{eq:BN}) becomes: $2 R_0-1=e^{2i\delta}$, which implies (\ref{eq:delta}). However, while with the general Bessho-Newman relation (\ref{eq:BN}) it is possible to obtain (\ref{eq:delta}) only in the domain $k<2\pi$, usage of the explicit forms (\ref{eq:A0}) and (\ref{eq:R0}), respectively for $\mathcal{A}_0^+$ and $R_0$, has allowed to extend (\ref{eq:delta}) to any wavenumber. Furthermore, by using (\ref{eq:T0}) and (\ref{eq:delta}), (\ref{eq:BN}) yields
\begin{equation}
\cos\delta=|\mathcal{A}_0^+|^{-1}\sum_{q=0}^{\bar{q}}\frac{\gamma_q}{\epsilon_q}\,\left|\mathcal{A}_q^+R_q\right|,
\label{eq:cosdelta}
\end{equation}
for any $k$. Expression (\ref{eq:cosdelta}) is a particular form of the Bessho-Newman relation, valid for a periodic array of flap-type converters under normally-incident waves. Note that for $k<2\pi$, i.e. $\bar{q}=0$, all the transverse modes are trapped near the array and (\ref{eq:cosdelta}) reduces to
\begin{equation}
\cos\delta=|R_0|,\quad k<2\pi.
\label{eq:cosdeltaR0}
\end{equation}
\subsection{Relation between $F$ and $\mathcal{A}_0^\pm$ (array Haskind relation)}
Consider the complex exciting torque (\ref{eq:F}) and the fundamental radiation coefficient (\ref{eq:A0}). Isolating the term $\alpha_{00}$ from (\ref{eq:A0}) and substituting it into (\ref{eq:F}), yields after some algebra
\begin{equation}
F=\pm2A_I\mathcal{A}_0^{\pm}C_g.
\label{eq:haskind}
\end{equation}
According to (\ref{eq:haskind}), the long-crested component of the radiated wave field is sufficient to obtain the exciting torque acting on each flap, for any value of $k$. Furthermore, since $A_I$ and $C_g$ are real numbers, (\ref{eq:haskind}) requires
\begin{equation}
F=|F|e^{i\delta}.
\label{eq:phaseF}
\end{equation}
Expression (\ref{eq:haskind}) can be transformed into physical variables via (\ref{eq:nondimvar}), thus giving
$$F'=2\rho g b'A_I'\mathcal{A}_0^+{'}C_g'.$$
The latter is similar in form to the well-known two-dimensional Haskind relation \cite{ME05} except for the factor $b'$, which represents the array spacing. Finally, note that (\ref{eq:haskind}) is an extension to intermediate water  depth of Srokosz's equation (4.3) in \cite{SR80}.
\subsection{Relation between $F$ and $R_0$}
The relation between the exciting torque and the fundamental reflection coefficient can be easily obtained by isolating $\beta_0$ from (\ref{eq:R0}), substituting it in (\ref{eq:F}) together with (\ref{eq:Zn}) and developing the algebra, so that
\begin{equation}
F=2A_IR_0\frac{\tanh kh}{k}\left(h-c+\frac{\cosh kc-\cosh kh}{k \sinh kh}\right).
\label{eq:relFR0}
\end{equation}
Hence the exciting torque acting on each flap is related to the amplitude of the long-crested component of the reflected wave field, for any value of $k$. 
\subsection{Relation between $R_0$ and $\mathcal{A}_0^+$}
\begin{figure}[t]
\begin{center}
\includegraphics[width=7.5cm, trim= 5cm 10cm 5cm 0cm]{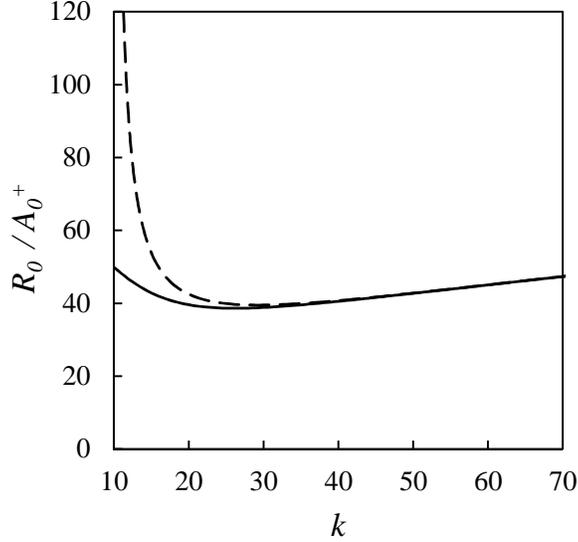}
\end{center}
\caption{Ratio $R_0/\mathcal{A}_0^+$ (\ref{eq:relR0A0p}) versus non-dimensional wavenumber $k$ (solid line) and approximate expression (\ref{eq:approxex}) for large $k$ (dashed line). Parameters of the system are $b'=91.6\,\mathrm{m}$, $h'=10.9\,\mathrm{m}$, $c'=1.5\,\mathrm{m}$.}
\label{fig:shortwav}
\end{figure}

By equating (\ref{eq:haskind}) and (\ref{eq:relFR0}) it is immediate to obtain
\begin{equation}
\frac{R_0}{\mathcal{A}_0^+}=\frac{kC_g}{\tanh kh \left(h-c+\frac{\cosh kc-\cosh kh}{k \sinh kh} \right)},
\label{eq:relR0A0p}
\end{equation}
valid for any $k$. Physically, (\ref{eq:relR0A0p}) measures the ratio between the reflective capacity of the system as an array of screens and the radiative capacity of the system as an array of wavemakers, oscillating at unison. In short waves, where the flaps are deemed to be operating \cite{HE10}, it is roughly $k\gg 1$, so that (\ref{eq:relR0A0p}) becomes
\begin{equation}
\frac{R_0}{\mathcal{A}_0^+}\simeq \frac{\sqrt{k}}{2\left(h-c-k^{-1}\right)},
\label{eq:approxex}
\end{equation}
as shown in figure \ref{fig:shortwav}.
Since the ratio (\ref{eq:approxex}) is $O(k^{1/2})$, the diffractive phenomena occurring in the system dominate over the radiative ones in short waves. This suggests that the effects of diffraction are not to be neglected if an accurate description of the system dynamics is to be pursued. 
\subsection{Relation between $\nu$ and $\mathcal{A}_0^+$}
Consider expression (\ref{eq:nu}), which defines the radiation damping $\nu$ for the reference plate. Isolating $\Im\left\lbrace\alpha_{00}\right\rbrace$ in (\ref{eq:A0}) and substituting it into (\ref{eq:nu}) yields, after some algebra,
\begin{equation}
\nu=2\Re\left\lbrace \mathcal{A}_0^+\right\rbrace \frac{\tanh kh}{k}\left(h-c+\frac{\cosh kc-\cosh kh}{k\sinh kh}\right),
\label{eq:relnua0}
\end{equation}
for any $k$. Incidentally, by isolating $\mathcal{A}_0^+$ in the array Haskind relation (\ref{eq:haskind}) and substituting it into (\ref{eq:relnua0}), the latter becomes
$$\nu=\Re\left\lbrace \frac{F}{A_I}\right\rbrace \frac{\tanh kh}{kC_g}\left(h-c+\frac{\cosh kc-\cosh kh}{k\sinh kh}\right),$$
which corresponds to expression (C3) of Renzi \& Dias \cite{RD12}. Note that (\ref{eq:relFR0}), (\ref{eq:relR0A0p}) and (\ref{eq:relnua0}) allow to obtain the exciting torque and the radiation damping - and consequently the optimum generated power (\ref{eq:Popt}) - directly from the fundamental reflection coefficient $R_0$. This is a peculiar property of the flap-type converter and does not hold in general for converters of different shape.

The above relations have been used to check the numerical calculations in this paper. In order to assess the accuracy of computations for a given equation of the form $\mathrm{l.h.s}=\mathrm{r.h.s.}$, the relative error 
\begin{equation}
\epsilon=\frac{|\mathrm{l.h.s.}-\mathrm{r.h.s.}|}{|\mathrm{r.h.s}|}
\label{eq:error}
\end{equation}
is defined. For a typical system configuration (see \ref{sec:appD}), taking 40 array modes, 5 depth modes and 5 terms in the Chebyshev expansion, is sufficient to obtain a relative error of $O (10^{-16})$ in calculating the Haskind relation (\ref{eq:haskind}) and $O(10^{-15})$ in calculating the remaining relations (\ref{eq:relFR0})--(\ref{eq:relnua0}). Hence the method of solution based on the Green's theorem of \cite{RD12} reveals to be fast convergent and very efficient. In the following, the influence of the array aperture on the performance of the system is assessed, based on the relations found in this section.

\section{Performance evaluation}
\label{sec:paran}
Consider the optimum capture factor
\begin{equation}
C_F^{opt}=\frac{1}{4}\frac{|F|^2}{\nu A_I^2 C_g w},
\label{eq:CFopt}
\end{equation}
obtained by substituting the optimum power output (\ref{eq:Popt}) into (\ref{eq:CF}). By replacing $F$ with (\ref{eq:relFR0}), $\nu$ with (\ref{eq:relnua0}), and by performing some algebra, (\ref{eq:CFopt}) can be rewritten as
\begin{equation}
C_F^{opt}=\frac{1}{2}\frac{|R_0|}{(1-a)\cos\delta},
\label{eq:CFopt2}
\end{equation} 
where $\delta$ is still the argument of $R_0$. According to (\ref{eq:CFopt2}), the performance of the array depends on the reflection coefficient magnitude and argument, which in turn are functions of the array aperture. Hence the solution of the scattering problem alone is sufficient to assess the performance of the system via (\ref{eq:CFopt2}). This result confirms that diffraction effects are fundamental in wave-power extraction from flap-type WECs. Therefore, the empiric criterion for which ``to absorb waves means to generate waves'' \cite{FA02}, valid for small floating bodies in the absence of diffraction, does not apply here in full. In the following, expression (\ref{eq:CFopt2}) will be validated against known theories in the small-gap and point-absorber limits. Then the maximum capture factor will be assessed.
\subsection{Small-gap limit}
In the limit $a\rightarrow0$, the flaps become joined to each other and the system is two-dimensional. In this case it is $R_0\rightarrow1$, because of complete reflection of the incident wave in the diffraction problem. Then it is straightforward to show that (\ref{eq:CFopt2}) becomes
\begin{equation}
C_F^{opt}\rightarrow\frac{1}{2},
\label{eq:CF2d}
\end{equation}
i.e. the capture factor coincides with the classical hydrodynamic efficiency for 2D devices \cite{ME05}.
\subsection{Point-absorber limit} 
\begin{figure}[t]
\begin{center}
\includegraphics[width=7cm, trim= 5cm 9cm 5cm 0cm]{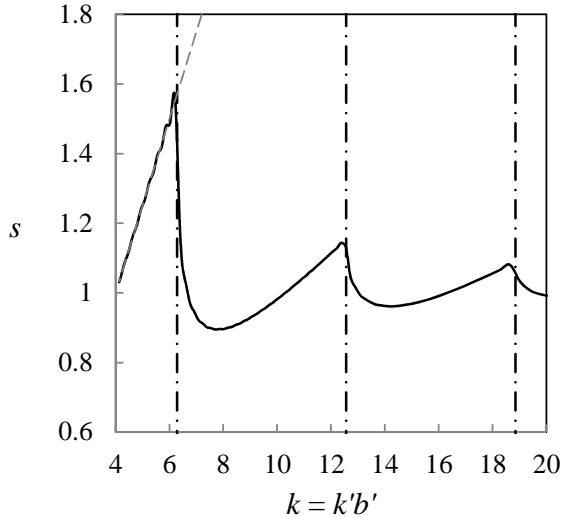}
\end{center}
\caption{Interaction factor (\ref{eq:qfactor}) versus non-dimensional wavelength $k$ for the point-absorber approximation. Parameters of the system are $b'=91.6\,\mathrm{m}$, $w'=4.58\,\mathrm{m}$, $h'=10.9\,\mathrm{m}$, $c'=1.5\,\mathrm{m}$. The ratio $w=w'/b'=0.05$ justifies the use of the point-absorber approximation. The vertical dash-dotted lines show the resonant wavenumbers $k=2(\bar{q}+1)\pi$; the grey dashed line shows the linear behaviour $s=k/4$ for $k<2\pi$.}
\label{fig:qfactor}
\end{figure}

Consider now the limit $w=w'/b'\rightarrow 0$, for fixed array spacing $b'$. In this limit, the wavelength $\lambda=2\pi/k$ of the incident wave is much larger than the flap width, $\lambda\gg w$, and the interaction between the flaps is weak. Hence the results of the present theory can be compared to those of Budal \cite{BU77} and Srokosz \cite{SR80} for an infinite array of point absorbers. For such a system, the efficiency is assessed via the absorption length
\begin{equation}
L'=\frac{P}{\frac{1}{2} A_I^2 C_g}\,b'=C_F w',
\label{eq:abslen}
\end{equation} 
which is the ratio between the power captured by the single device and that incident per unit wave crest length. For an array of converters the optimum absorption length is
\begin{equation}
L'_{opt}=l'_{opt}\,s,
\label{eq:abslen2}
\end{equation}
where $l'_{opt}$ is the optimum absorption length for an isolated body and $s$ is an interaction factor \cite{BU77, SR80}. When $w\ll \lambda $, each flap can be considered as a three-dimensional axisymmetric body, whose optimum absorption length for given wavelength of the incident wave is 
\begin{equation}
l'_{opt}=\xi \frac{\lambda}{2\pi}\,b',
\label{eq:pointabs}
\end{equation}
where $\xi=1$ for heave and $\xi=2$ for surge (see \cite{EV76,HE10}). By substituting (\ref{eq:pointabs})  into (\ref{eq:abslen2}), then the latter into (\ref{eq:abslen}) and employing (\ref{eq:CFopt2}) for the optimum capture factor, the interaction factor becomes
\begin{equation}
s=\frac{k}{2\xi}\frac{|R_0|}{\cos\delta},
\label{eq:qfactor}
\end{equation}
where $\xi=2$, because in the point-absorber approximation the flap moves essentially in surge \cite{HE10}. Figure \ref{fig:qfactor} shows the plot of $s$ versus the non-dimensional wavenumber $k=k'b'$ for a typical configuration in which $w=0.05$. 
When $k< 2\pi$ (i.e. $\lambda> 1$) all the transverse modes are trapped and the behaviour of the curve is linear: $s\simeq k'b'/(2\xi)$. This agrees formally with the results shown by Budal \cite{BU77} and Srokosz \cite{SR80} for an infinite array of heaving point-absorbers, where $\xi=1$. When $k>2\pi$, i.e. $\lambda<1$, incomplete trapping of the transverse modes strongly modifies the behaviour of the curve. The interaction factor globally decreases, but spikes occur near the resonant wavenumbers $k=2(\bar{q}+1)\pi$. Note that this dynamics is different from that shown in \cite{SR80} for a system of heaving point absorbers. In the latter, $s$ drops to zero when $k\rightarrow2(\bar{q}+1)\pi$ from the left, so that the trapping of the transverse modes has a detrimental effect on the performance of the system \cite[see fig. 2]{SR80}. Here, instead, resonance produces local maxima of $s$ near $k=2(\bar{q}+1)\pi$ (see again figure \ref{fig:qfactor}) and therefore is beneficial in increasing the optimum efficiency, even away from complete trapping. This happens because the resonance of the transverse modes enhances the horizontal (surge) actions and drops the vertical (heave) loads \cite{CH94}. Therefore surging WECs benefit the most from the resonant mechanisms activating in an array configuration.
\subsection{Maximum capture factor}
Consider again the optimum capture factor, given by (\ref{eq:CFopt2}). Expression (\ref{eq:cosdelta}) shows that $|R_0|/\cos\delta$ has a unit upper limit (\ref{eq:cosdeltaR0}) when all the transverse modes are trapped near the array. This situation is the most favourable for energy extraction and yields the maximum capture factor attainable by an array of oscillating wave energy converters of given aperture $a$. Substitution of (\ref{eq:cosdeltaR0}) into (\ref{eq:CFopt2}) yields
\begin{equation}
C_F^{opt}= C_F^{max}=\frac{1}{2(1-a)},\quad k<2\pi
\label{eq:maxCF}
\end{equation}
for the maximum capture factor. Incidentally, note that (\ref{eq:maxCF}) correspond to Srokosz's \cite{SR80} maximum efficiency $E^{max}=wC_F^{max}$ for a symmetric floating body in a channel. Since $a<1$, the maximum capture factor (\ref{eq:maxCF}) for the array configuration is larger than the well-known limit value of $1/2$, obtained in the small-gap approximation $a\rightarrow 0$ (see expression \ref{eq:CF2d}). Hence the mutual interaction between the flaps, which is responsible for the trapping of energy near the array in the form of short-crested waves, can increase the capture factor of the system \cite{RD12}. 

\section{Application to wave energy extraction}
\label{sec:waven}
In applications, the incident wave period $T'$ and wavelength $\lambda'$ are known, together with the flap width $w'$. The array aperture $a$ needs to be optimised so that the capture factor is maximum. Expression (\ref{eq:maxCF}) for the maximum capture factor would suggest to increase the aperture as much as possible, so that $a\rightarrow 1$ and consequently $C_F^{max}\rightarrow \infty$. However, (\ref{eq:maxCF}) is to be regarded as a theoretical upper limit. This is motivated by two reasons. First and most important, as $a\rightarrow 1$, then $w=1-a\rightarrow 0$. Now recall that $w=w'/b'$, being $w'$ the width of the single flap, which in practical applications is large. As a consequence, in order to have $w\rightarrow 0$, it must be $b'\rightarrow \infty$. In this limit, $\lambda=\lambda'/b'\rightarrow 0$, $k\rightarrow \infty$ and expression (\ref{eq:maxCF}) for the maximum capture factor is no longer valid. Physically, by increasing the array aperture $a$, the spatial period $b'$ increases so much, that the array is no longer able to trap all the transverse modes, resulting in more energy leakage. Second, recall that expression (\ref{eq:maxCF}) for the theoretical maximum capture factor is obtained under the assumption (\ref{eq:bodyres}), i.e. that the self-oscillation frequency of each flap is tuned to the frequency of the incoming waves. However, flap-type converters are usually designed to avoid this eventuality. At body resonance, the stroke of the flap would exceed by far the amplitude of the incident wave \cite{HE10}. It is then clear that this condition is undesirable and not compatible with the power take-off mechanism \cite{CR08}. Away from body resonance, the maximum values of $C_F$ attained are smaller than those predicted by (\ref{eq:maxCF}). This is due to the presence of the inertial terms at the denominator of $P$ (\ref{eq:pow}), which in turn reduce $C_F$ (\ref{eq:CF}).

The mathematical model of \S\ref{sec:model} is now applied to determine an optimisation criterion for the array aperture $a$, which maximises the power output of an array of flap-type WECs. The configuration investigated here is that of an infinite array of converters similar to Oyster 800\texttrademark\footnote{Oyster is a trademark of Aquamarine Power Limited.}. Each converter has a width $w'=26\,\mathrm{m}$ and is placed upon a foundation of height $c'=1.5\,\mathrm{m}$ from the bottom of the ocean; water depth is $h'=10.9\,\mathrm{m}$. Monochromatic incident waves of amplitude $A'_I=1\mathrm{m}$, period $T'=7\,\mathrm{s}$ (wavelength $\lambda'= 62\,\mathrm{m}$), representative on average of the wave climate off the west coast of Ireland \cite{R12}, are considered. For these parameters, several different layouts, from compact ($a=0.30$) to sparse ($a=0.95$), are analysed to determine the optimum array aperture $a_{opt}$. In each case the power $P'=\rho A{'}^2b{'}^{3/2}g^{3/2}P$ extracted by a single flap and the relevant capture factor $C_F$ (\ref{eq:CF}) are calculated with the mathematical model of \S\ref{sec:model}. Table \ref{tab:1} shows the selected values of $a$, the corresponding value of the array spacing $b'$, the generated power $P'$ in $\mathrm{kW}$, the capture factor $C_F$ and the theoretical maximum $C_F^{max}$ . 
\begin{table}\centering
\begin{tabular}{c c c c c c c}
\hline
 $a$ & $0.3$& $0.40$ & $0.50$ & $a_{opt}=0.58$ & $0.70$ & $0.95$ \\ \hline
 $b'\,(\mathrm{m})$ & 37 & 43 & 52 & 62 &87 & 520\\\hline
$P'\,(\mathrm{kW})$ & 504 & 564 & 660 & 795 & 574 & 605\\ \hline
$C_F$ & 0.60 & 0.68 & 0.79 & 0.95 & 0.69 & 0.73\\ \hline
$C_F^{max}$ & 0.71 & 0.83 & 1 & 1.19 & (1.67) & (10)\\\hline
\end{tabular}
\caption{Array spatial period $b'$, power output $P'$ , capture factor $C_F$  and maximum theoretical capture factor $C_F^{max}$ for an infinite array of flap-type converters similar to Oyster 800\texttrademark. Different apertures are considered, from compact ($a=0.3$) to sparse ($a=0.95$). Calculations are made with the mathematical model of \S\ref{sec:model}.}
\label{tab:1}
\end{table}
The largest power output and capture factor are attained at the optimum configuration $a=a_{opt}=0.58$, which corresponds to $\lambda'=b'$ (i.e. $k=2\pi$), the trapping wavelength of the first transverse mode \cite{RD12}. For $a<a_{opt}$, $\lambda'>b'$ (i.e. $k<2\pi$) and all the transverse modes are perfectly trapped (see table \ref{tab:1}). However, since $a$ is small, the theoretical maximum (\ref{eq:maxCF}) sets a relatively small upper limit for $C_F$. By increasing $a$, $C_F^{max}$ increases and so does the actual capture factor $C_F$, until it reaches its maximum at $a=a_{opt}$. For $a>a_{opt}$ the theoretical limit $C_F^{max}$ still increases, while the actual capture factor $C_F$ decreases. This happens since in these cases $\lambda'<b'$ (i.e. $k>2\pi$) and complete trapping of the transverse modes is not possible, so that (\ref{eq:maxCF}) does not hold in practice. Energy leakage associated to the propagating transverse waves lowers the power absorption of the array well below the theoretical maximum values. In conclusion, the optimum aperture that maximises the capture factor is the one for which $\lambda'=b'$, i.e.
\begin{equation}
a_{opt}=1-\frac{w'}{\lambda'},
\end{equation}  
which can be used as a preliminary design formula.

The theory exposed here reveals to be useful for the optimisation of the efficiency of an infinite array of flap-type wave energy converters in incident monochromatic waves of given period. Of course, in real seas superposition of different wave components must be considered. The power output may thus vary, depending on the coupling between the spectrum energy period and the torque peak period \cite{R12, CL12}. Further analysis is therefore necessary to obtain more accurate estimates of wave power generation in random seas. Finally, due to real sea bottom conditions, converters in array are likely to be deployed in a staggered configuration and in a finite number. Ongoing work is investigating the dynamics of a finite array of staggered converters and will be disclosed in the near future.
\section{Conclusions}
A periodic array of flap-type WECs has been analysed in this work by using the semi-analytical model of Renzi \& Dias \cite{RD12}. Asymptotic analysis in the far field has allowed to obtain new expressions for the radiation, reflection and transmission coefficients. Relations have been determined between the $0$th-mode coefficients and the hydrodynamic parameters of the system. Some of these relations constitute an extension to intermediate water of the previous results obtained by Srokosz \cite{SR80} for an array of floating bodies in deep water, while some others are peculiar to the flap-type converter. The efficiency of the system, evaluated via the capture factor, has been shown to depend on the reflection coefficient magnitude and argument, which in turn are functions of the array aperture. This result shows that diffraction effects are fundamental in wave-power extraction from flap-type WECs. Unlike a line of heaving buoys \cite{SR80}, an array of flap-type WECs can exploit the resonance of transverse modes to attain high capture factor levels, even when complete trapping of the transverse modes does not occur. The maximum capture factor is attained in the regime of complete trapping, for which the amount of energy available for extraction is the largest. Given the wave period and the flap width, the capture factor can be maximised by varying the spacing between the flaps, such that complete trapping of the transverse modes occurs. These results have been obtained under the assumptions that the fluid is inviscid and the flow is irrotational. Viscous effects and turbulent dissipations may reduce the values predicted here, especially near trapping frequencies (see \cite{CH94}).  \\

This work was funded by Science Foundation Ireland (SFI) under the research project ``High-end computational modelling for wave energy systems''. Discussions with Prof. D.V. Evans and Dr X. B. Chen have been illuminating. Numerical data provision by Dr G. Bellotti and Mr A. Abdolali is gratefully acknowledged.

\appendix
\section{Asymptotic analysis of a summation}
Consider the sum
\begin{equation}
S(X,Y)=\sum_{m=-\infty}^{+\infty}\frac{H_1^{(1)}\left(k\sqrt{X^2+(Y-m)^2}\right)}{\sqrt{X^2+(Y-m)^2}},
\label{eq:S}
\end{equation}
where $k$ is a real positive number. To determine the asymptotic behaviour of $S$ for $|X|\rightarrow\infty$, first consider expressions (2.7) and (2.13) of \cite{L98}, which together give
$$\sum_{m=-\infty}^{\infty} H_0^{(1)}\left(k\sqrt{X^2+(Y-m)^2}\right)=\frac{2}{i}\sum_{q=-\infty}^{+\infty}\frac{e^{-\sqrt{(2q\pi)^2-k^2}\,|X|}}{\sqrt{(2q\pi)^2-k^2}}\,e^{2iq\pi Y}.$$ Differentiating the latter by using the property $H_0^{(1)}{'}=-H_{1}^{(1)}$ and substituting the result in (\ref{eq:S}) yields, after some elementary manipulations,
\begin{equation}
S(X,Y)=\frac{2}{ikX}\,\mathrm{sign}(X) \sum_{q=-\infty}^{+\infty} e^{i\gamma_q k|X|} e^{2iq\pi Y},
\label{eq:S2}
\end{equation}
where
\begin{equation}
\gamma_q=\sqrt{1-(2q\pi/k)^2}.
\label{eq:gammaq}
\end{equation}
Now note that the argument of the square root in $\gamma_q$ (\ref{eq:gammaq}) is positive only if $|q|<k/(2\pi)$. Then define $\bar{q}$ as the largest integer for which
\begin{equation}
|\bar{q}|<\frac{k}{2\pi}.
\end{equation}
For $|q|<|\bar{q}|$, $\gamma_q$ is real and the relevant terms in $S$ (\ref{eq:S2}) are oscillating functions of $X$. On the other hand, for $|q|>|\bar{q}|$, $\gamma_q$ is purely imaginary and the relevant terms in $S$ decay exponentially with $X$. Hence neglecting the evanescent terms in (\ref{eq:S2}), transforming the exponential in $Y$ with the Euler formula and developing some straightforward algebra yields
\begin{eqnarray}
S(X,Y)&=&\sum_{m=-\infty}^{+\infty}\frac{H_1^{(1)}\left(k\sqrt{X^2+(Y-m)^2}\right)}{\sqrt{X^2+(Y-m)^2}}\sim  \frac{2}{ikX}\,\mathrm{sign}(X)\nonumber\\
&\times&\sum_{q=0}^{\bar{q}}\epsilon_q e^{i\gamma_q k|X|}\cos(2q\pi Y),\quad |X|\rightarrow+\infty,
\label{eq:Sff}
\end{eqnarray} 
where $\epsilon_q=2-\delta_{0q}$ is the Jacobi symbol and $\delta_{nm}$ the Kronecker symbol, $n,m\in\mathbb{N}$.
\section{Evaluation of an integral}
Consider the integral
\begin{equation}
I_{pq}=\int_{-1}^{1}(1-u^2)^{1/2}U_{2p}(u)\cos\left[2q\pi\left(y-\frac{wu}{2}\right)\right]\,du,
\label{eq:Ipq}
\end{equation}
where $p$ and $q$ are integers and $|y|<1/2$. For $q=0$, application of the property (7.343) of \cite{GR07} for the Chebyshev polynomials $U_{2p}$ gives immediately
\begin{equation}
I_{p0}=\int_{-1}^{1}(1-u^2)^{1/2}U_{2p}(u)\,du=\frac{\pi}{2}\,\delta_{p0}.
\label{eq:Ip0}
\end{equation}
Now consider the case $q>0$. Expanding the cosine in (\ref{eq:Ipq}), performing the substitution $u=\cos\theta$ and using the property 
$$U_{2p}(\cos\theta) = \frac{\sin [(2p+1)\theta]}{\sin\theta},$$
yields after some algebra
\begin{equation}
I_{pq}=\cos(2q\pi y)\int_{0}^{\pi/2} 2\sin\theta\sin [(2p+1)\theta] \cos(z\cos\theta)\, d\theta,
\label{eq:I2}
\end{equation}
where $z=q\pi w$. Substituting the identity $$2\sin\theta\sin [(2p+1)\theta]=\cos(2p\theta)-\cos[(2p+2)\theta]$$ in (\ref{eq:I2}), using the property 
$$\int_{0}^{\pi/2}\cos 2p\theta \cos(z\cos \theta)\,d\theta=(-1)^p\frac{\pi}{2}J_{2p}(z)$$
\citep[see \S3.714 of][]{GR07}, where $J_{2p}$ is the Bessel function of first kind and order $2p$, and going back to the original variables yields finally
\begin{eqnarray}
I_{pq}&=&\int_{-1}^{1}(1-u^2)^{1/2}U_{2p}(u)\cos\left[2q\pi\left(y-\frac{wu}{2}\right)\right]\,du=\cos (2q\pi y)(-1)^p\nonumber\\
&\times& (2p+1)\frac{J_{2p+1}(q\pi w)}{qw}, \quad q>0,
\label{eq:Ipq2} 
\end{eqnarray}
where the relation $J_{n-1}(z)+J_{n+1}(z)=2n J_n(z)/z$ \citep[see \S 8.471 of][]{GR07} has also been used.
\section{Comparison with numerical model}
\label{sec:appD}
\begin{figure}[t]
\begin{center}
\includegraphics[width=7.5cm, trim= 8cm 12.5cm 7cm 1cm]{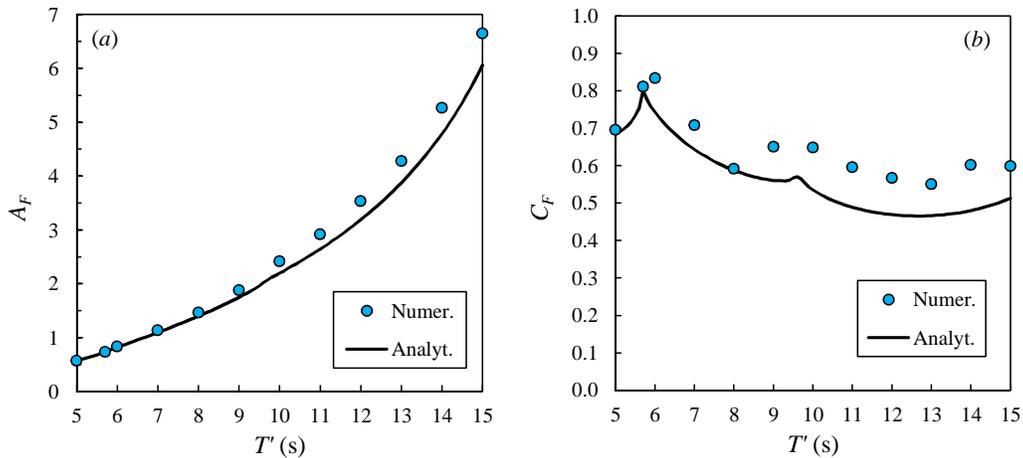}
\end{center}
\caption{($a$) Amplification factor vs period of the incident waves. ($b$) Capture factor vs period of the incident waves. Parameters of the system are $b'=91.6\,\mathrm{m}$, $w'=18\,\mathrm{m}$, $h'=10.9\,\mathrm{m}$, $c'=1.5\,\mathrm{m}$. The solid line shows the results obtained with the analytical model of Section \ref{sec:model}, dots show the results of the numerical model.}
\label{fig:power}
\end{figure}

In this section the mathematical model of \S\ref{sec:model} is further validated against available numerical results. The latter have been obtained with a finite-element numerical model developed by the University of Roma Tre (Italy), as detailed in \cite{Retal12}. In the numerical model, the array layout is replaced by the equivalent configuration of a single plate of width $w'/2$ on the side of a channel of width $b'/2$. This allows to speed up the calculations without loss of physical meaning. The numerical model solves the equation of motion (\ref{eq:lapl}), with boundary conditions (\ref{eq:bcsurf}) on the free-surface, (\ref{eq:bottom}) on the bottom, (\ref{eq:simm}) on the channel lateral walls and (\ref{eq:plate}) on the flap. A radiation condition, including a source term for generating the desired incoming waves, is imposed at the open generation boundary, which also allows the waves reflected back by the device to  leave the computational domain freely. At the end of the flume, a sink term is imposed to simulate an open boundary, where the transmitted waves leave the domain freely. The flume length is 3 times the wave length to assure that there is enough space for waves to develop and then leave the domain. In the geometry chosen for comparison, each flap has width $w'=18\,\mathrm{m}$ and the ocean has depth $h'=10.9\,\mathrm{m}$. The foundation is $c'=1.5\,\mathrm{m}$ tall and the spatial period of the array is $b'=91.6\,\mathrm{m}$, which corresponds to an array aperture $a\simeq 0.8$. The flap thickness is null in the semi-analytical model and equal to $1.8\,\mathrm{m}$ in the numerical model. Comparisons between the semi-analytical model of \S\ref{sec:model} and the numerical results are shown in figure \ref{fig:power}.
In the left panel the amplitude factor $A_F$ (\ref{eq:AF}) is plotted versus the period of the incident waves for the geometry described above. Agreement between analytical and numerical data is very satisfactory. Overall, $A_F\geq 0.5$ for all periods considered, meaning that the flaps effectively convert the wave motion into pitching motion. Figure \ref{fig:power} (right panel) shows the behaviour of the capture factor $C_F$ (\ref{eq:CF}) versus the incident wave period. Comparison between analytical and numerical data is very good at small periods, while the numerical model predicts larger values with longer waves. This is likely to be a thickness-induced effect, which becomes important with larger oscillations of the flap at larger periods. However, even at large $T'$ the results predicted by the two models are still in satisfactory general agreement. Note also that the capture factor is $C_F\geq 0.6$ in the interval $T\in[5,8]\,\mathrm{s}$, indicating that flap-type WECs are most effective in short waves.



\bibliographystyle{model3-num-names}
\bibliography{<your-bib-database>}



\end{document}